\documentclass[journal, 10pt]{IEEEtran}
\pdfoutput=1

\usepackage{times}
\usepackage{amsfonts, amssymb, amsmath,url}
\usepackage{graphicx}
\usepackage{multicol}
\usepackage[center]{caption}

\newcommand{\celsius}{$^\circ$C }
\renewcommand{\baselinestretch}{0.95}

\begin{document}
%
\title{Three Experiments to Analyze the Nature of the Heat Spreader}
%
%
%

\author{Seema Sethia$^1$, Shouri Chatterjee$^2$, Sunil Kale$^3$, Amit Gupta$^4$, Smruti R. Sarangi$^5$ \\
$^{1,2}$ Department of Electrical Engineering, IIT Delhi \\
$^{3,4}$ Department of Mechanical Engineering, IIT Delhi \\
$^5$ Department of Computer Science and Engineering, IIT Delhi \\
ersima.27@gmail.com, \{shouri@ee, srk@mech, agupta@mech, srsarangi@cse\}.iitd.ac.in
}

\maketitle

\begin{abstract}
In this paper, we describe ongoing work to investigate the properties of the
heat spreader, and its implication on architecture research. In specific, we
conduct two experiments to quantify the heat distribution across the surface of a
spreader during normal operation. The first experiment uses T-type thermocouples, 
to find the temperature difference across different points on the spreader. We observe
about 6\celsius difference on average. In the second experiment, we try to 
capture the temperature gradients using an infrared camera. However, this experiment
was inconclusive because of some practical constraints such as the low emissivity
of the spreader. We conclude that to properly model the spreader, it is necessary
to conduct detailed finite element simulations. We describe a method to accurately
measure the thermal conductivity of the heat spreader such that it can be used
to compute the steady state temperature distribution across the spreader.
\end{abstract}

\section{Introduction}
\begin{figure}[!htb]
\begin{center}
\includegraphics[width=0.8\columnwidth]{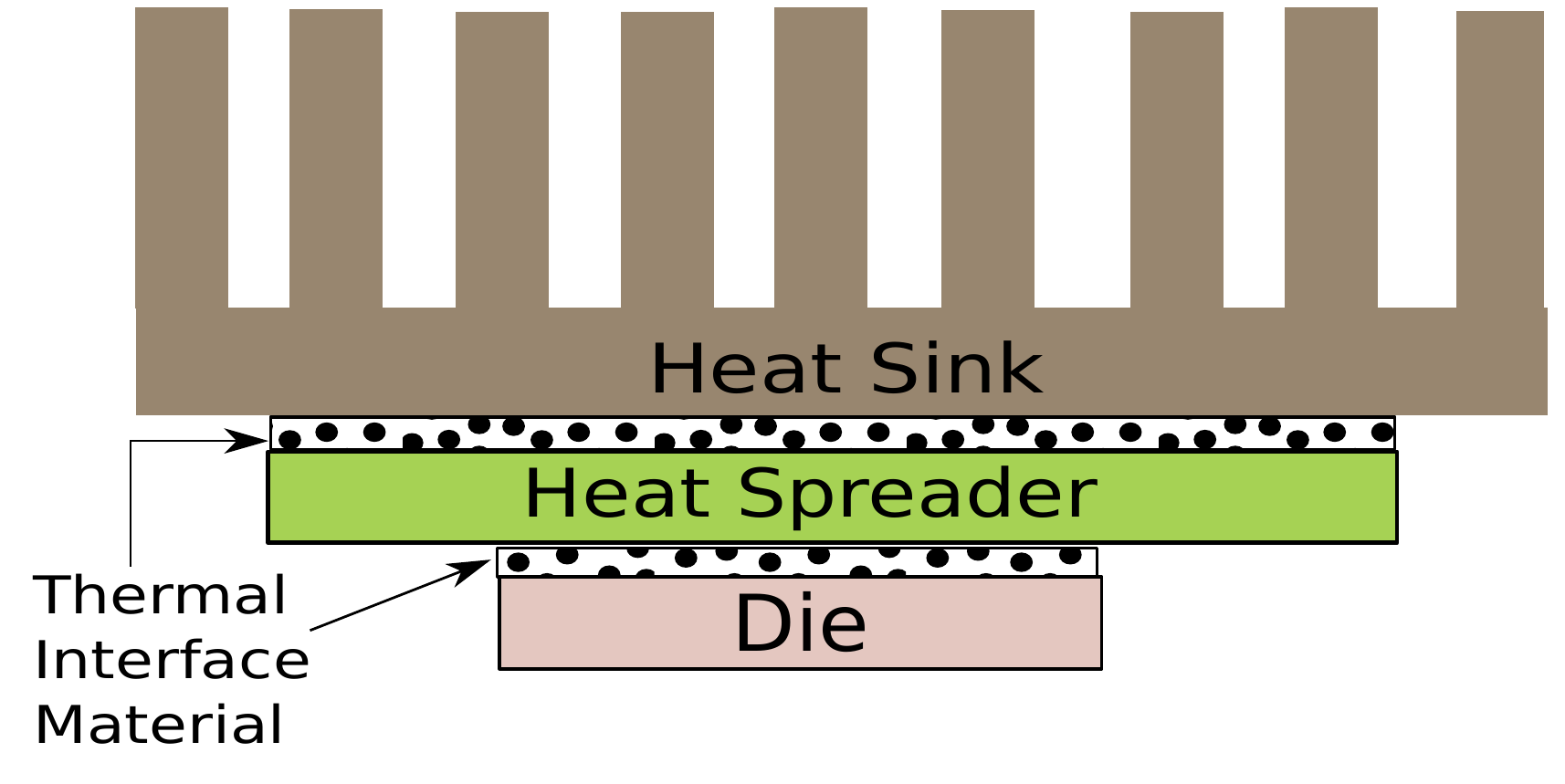}
\caption{ The chip package \label{fig:spreader}}
\end{center}
\end{figure}

An oft-ignored aspect of architecture level thermal modeling
is the heat spreader. The {\em heat spreader} is typically a nickel
coated copper plate placed between the die and the heat sink (see Figure~\ref{fig:spreader}).  
Its main role is to uniformly dissipate the heat generated by the die, and
transmit the heat to the heat sink. The heat sink is a large fin based heat exchanger
that is used to effectively dissipate the heat to the surrounding
air. 
The heat spreader effectively ``spreads out'' the heat and
reduces the severity and incidence of thermal hot spots. 

Given the fact that the heat spreader is nothing more than a metal plate, and does
not have a lot of inherent complexity, it has not received a lot of attention
by the architecture community. Some prior work such as~\cite{eval,wddd}, have treated
it as an isotherm (equal temperature at all points). We experimentally disprove
this hypothesis in this paper. Skadron et. al.~\cite{hotspot} 
treat the spreader as a mesh of points, where each point is a heat source and two adjacent
points are connected by a thermal resistance in their widely available thermal
modeling tool, HotSpot. However, there has been some recent criticism
of the equivalent thermal circuit based approach adopted by HotSpot in~\cite{images,comp}.
These works have reported a mean error of about 10\% in HotSpot.

We are currently working on developing a new temperature estimation tool. During the course
of this work, we wish to look at the spreader from an experimental viewpoint. 
There are several reasons for our belief that the spreader warrants a more thorough study. 
(1) The conductivity of silicon is roughly 100 W/m-K~\cite{comp}, whereas the conductivity
of the spreader is about 400 W/m-K. Consequently, the spreader is a far more efficient
lateral conductor of heat than silicon especially at distances of the order
of the dimensions of the die. This has important
implications for floor planning, thermal management, and task allocation in multicores. 
For example, it is possible for a set of active cores to heat up a set of relatively
quiescent cores by passing heat through the spreader. This will hurt  the 
performance of the relatively inactive cores, as well as long term lifetime reliability. 
(2) We can use the spreader temperature data that we collect to calibrate temperature simulators.
(3) We can measure some thermal properties of the spreader and use it to quantify the degradation
of the material over time. We can use this empirical data to perform more accurate 
FEM simulations. Lastly, for Intel based chips that integrate the spreader with the die,
it is not possible to study the temperature profile of the die independently. We need
to infer its thermal profile by analyzing the temperature gradients on the spreader.

\begin{figure*}[!htb]
\begin{center}

\begin{tabular}{cc}

\begin{minipage}{0.45\textwidth}
\includegraphics[width=0.7\columnwidth]{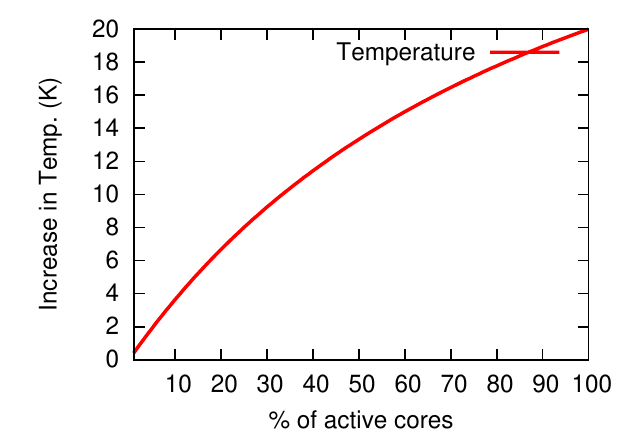}
\caption{Mean temp. of the inactive cores \label{fig:active} }
\end{minipage}
&
\begin{minipage}{0.45\textwidth}
\includegraphics[width=0.7\columnwidth]{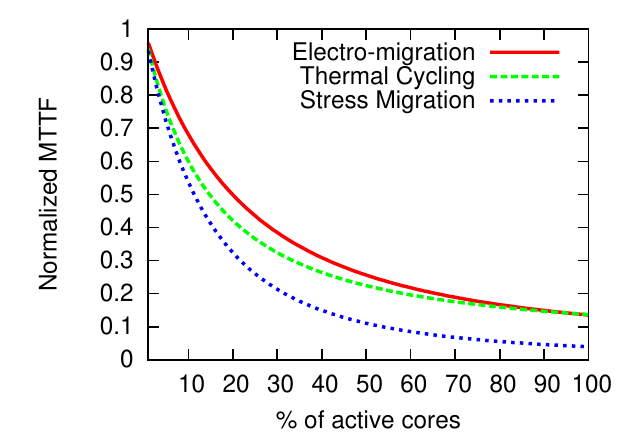}
\caption{MTTF of the inactive cores \label{fig:mttf} }
\end{minipage}
\\
\end{tabular}

\end{center}
\end{figure*}

We performed a simple thought experiment as follows. We consider a die with a large
number of cores (128), and assumed that there is lateral heat conduction just through
the spreader. We use typical parameters from the HotSpot tool (version 5.0)~\cite{hotspot}, and
simulated a scenario in which each core dissipates enough power to increase the
die temperature measured at the center by 20\celsius. We now turn off a set of cores, 
and measure the effect that the active cores have on the inactive cores. We 
plot the average temperature rise of the inactive cores in Figure~\ref{fig:active} 
 as a fraction of the number of active cores.
Figure~\ref{fig:mttf} shows the normalized decrease in the MTTF (Mean Time to
Failure) for three major failure
mechanisms: Electro-migration, Thermal Cycling, and Stress Migration (see
Srinivasan et. al.~\cite{jayanth}). 
We observe that lateral heat conduction can have a significant impact on inactive cores.
It can heat them up by 10 to 20\celsius, and can decrease their MTTF by upto 10X. 

In this paper, we describe two approaches to measure the temperature
distribution across the spreader during normal operation. We can
use these numbers to calibrate temperature estimation tools. Lastly,
we describe an approach to compute the thermal conductivity of the
spreader such that it can be used to seed FEM simulations.

\section{Experiments}

For all our experiments we use a 775 pin, 90 nm,  Intel Pentium 4(Prescott) chip mounted
on a Dell 00M075 Dimension 4300 Motherboard. It has a nominal frequency
of 3.06GHz, 1MB L2 cache, and has two voltage
steppings -- 1.25 and 1.388V. We perform three experiments: 
\begin{enumerate}
\item Measure the temperature distribution on the surface of the
 heat spreader using thermocouples.
\item Capture the temperature distribution with an infrared camera. 
\item Measure the thermal conductivity of the heat spreader using thermocouples
for accurate steady state
FEM simulation. 
\end{enumerate} 

\subsection{Design of Thermocouples}
A thermocouple consists of two wires made of different metals/alloys. At the point of
contact with the target material an EMF(voltage differential)
is generated between the two wires because of
their differing thermal properties. This difference in voltage is typically proportional
to the temperature of the target, and can be detected with a simple electronic circuit. 
Due to their simplicity and accuracy, they are commonly used to perform accurate temperature
measurements. In our experiments we use 300 $\mu$m T-type thermocouples made of copper and
constantan wires. They operate best in a temperature range between $\pm 200^\circ$C. 
To eliminate effects of corrosion and ageing, we used brand new wires. Secondly, to ensure
good contact between the wires, we heat the tip of both wires such that they fuse
together to make a strong junction. 

We connected the thermocouples to an Expert EX9018P data acquisition module. This is a
sophisticated analog to digital converter that converts the thermocouple voltages
to digital signals. It can process upto 8 thermocouple inputs. It has an internal multiplexer
that chooses one of them. The final output is in the RS 485 serial bus format. We subsequently
use an Advantech ADAM 4520 converter to convert the RS 485 signals to RS 232 signals that
can directly be fed to the serial port of a  standard PC. The ADAM 4520 chip also helps
to isolate the PC from ground loops and destructive voltage spikes.  We calibrated
the thermocouples with distilled boiling water and ice. New Delhi is 216 meters above sea level. 
The boiling point at this altitude is 99.304$^\circ$C  for a typical atmospheric pressure
of 987.56 millibars.

\begin{figure*}[!htb]
\begin{center}
\begin{tabular}{ccc}

\begin{minipage}{0.45\textwidth}
\includegraphics[width=0.99\columnwidth]{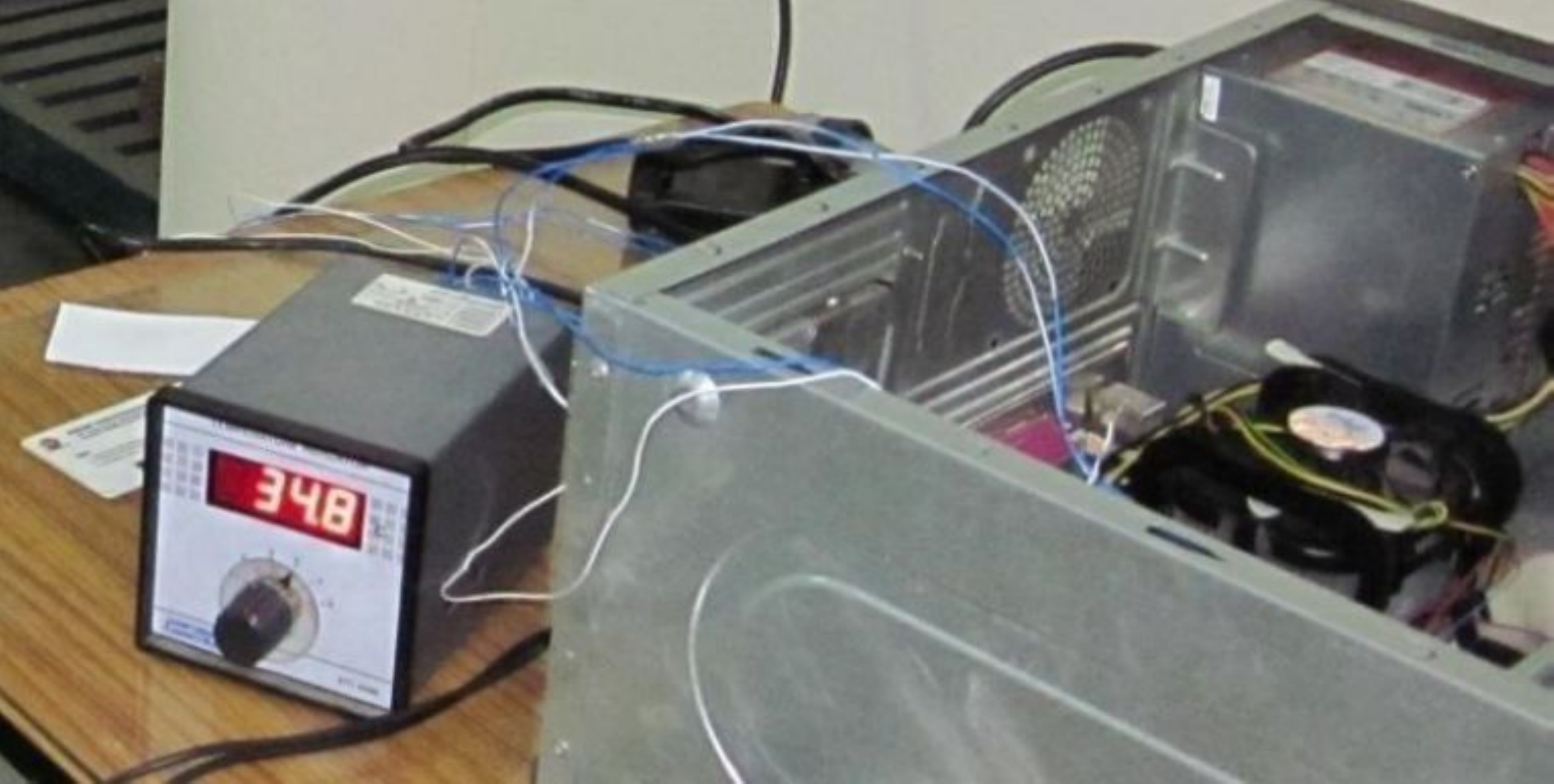}
\caption{Experimental setup \label{fig:setup}}
\end{minipage}

&

\begin{minipage}{0.23\textwidth}
\includegraphics[width=0.99\columnwidth]{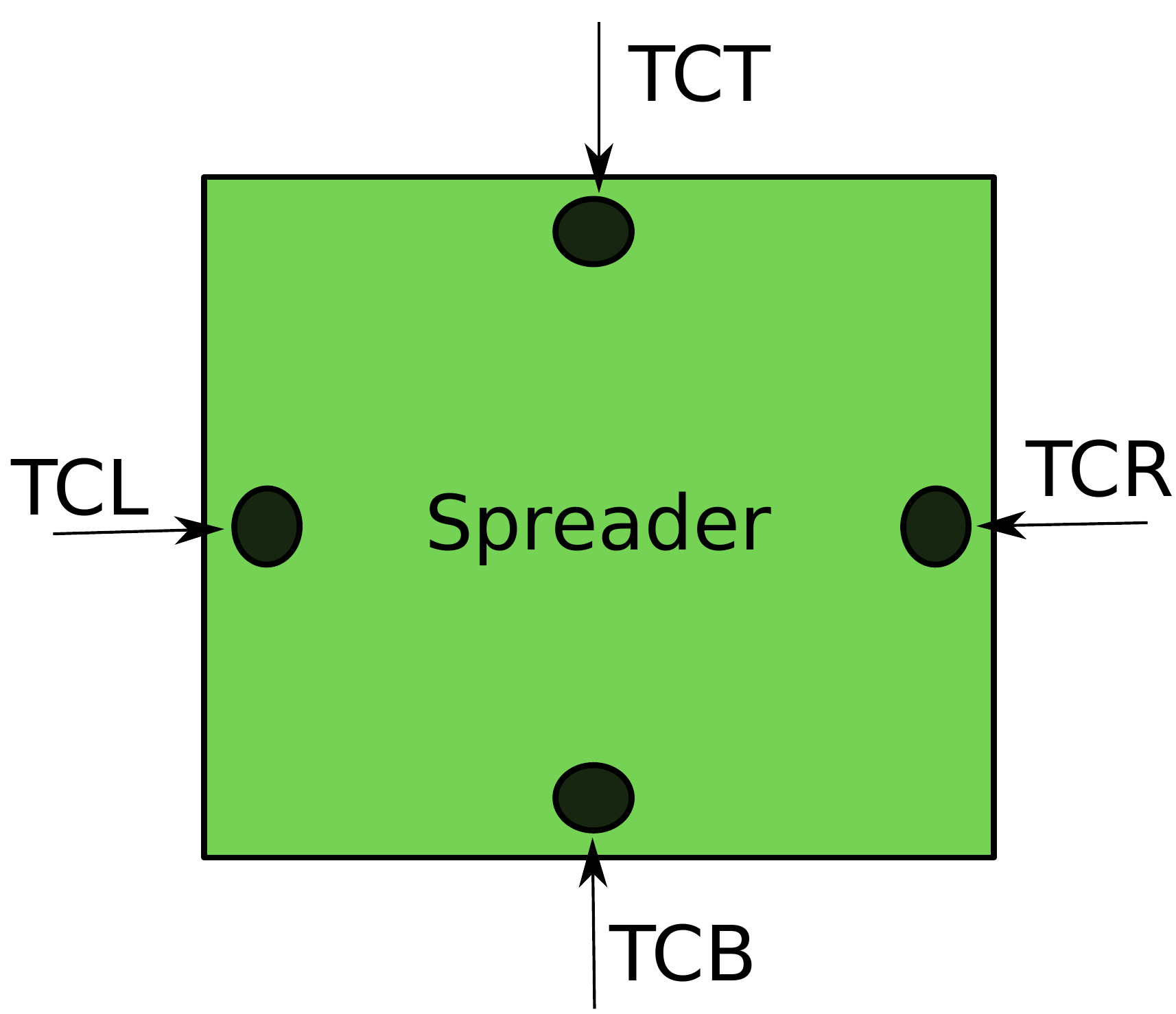}
\caption{Position of thermocouples on the spreader \label{fig:sensors}}
\end{minipage}

&
\begin{minipage}{0.22\textwidth}
\includegraphics[width=0.99\columnwidth]{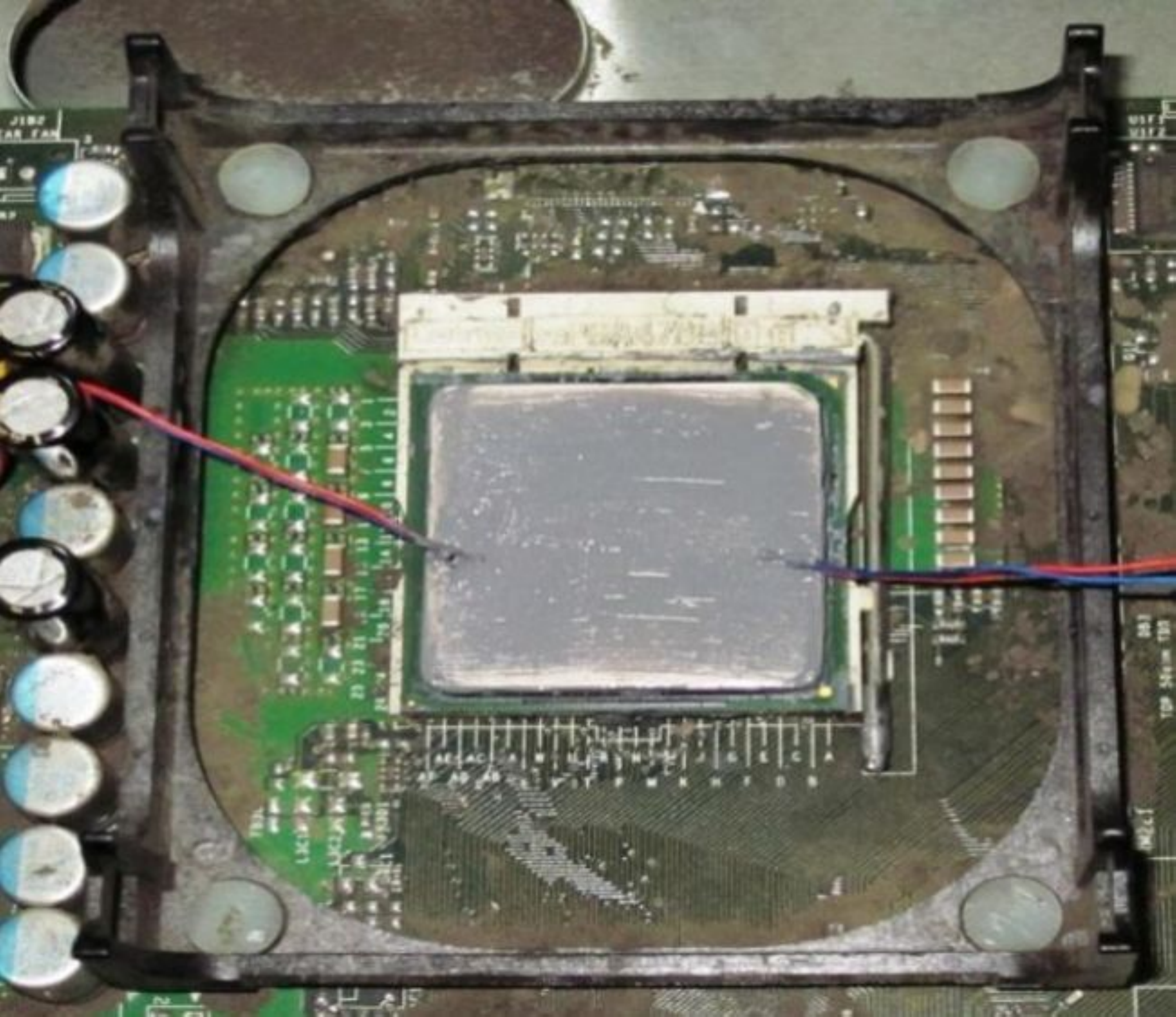}
\caption{Two attached thermocouples \label{fig:position}}
\end{minipage}
 
\\
\end{tabular}
\end{center}
\end{figure*}

\subsection{Experiment I - Thermocouple based Measurement}
\label{sec:exp1}
In this experiment we place thermocouples at different ends of the integrated heat spreader. 
We mostly follow the reference procedure as described in the Intel Thermal Design Guidelines
Document~\cite{intel-therm} (Appendix D). However, instead of applying the Kapton adhesive,
we apply Halnziye HY 610 thermal paste to achieve the dual purpose of making the thermocouples
stick to the spreader and provide good thermal conductivity for accurate measurement. This
thermal paste has mild adhesive properties.
Figure~\ref{fig:setup} shows our setup. Secondly, it was not necessary to drill holes through the
heat sink since we do not connect the thermocouples at the center. We attach them to the middle
of the four sides as shown in Figure~\ref{fig:sensors}. We define four positions on the spreader -- TCT, TCL, TCB, and
TCR.
We allow the setup to reach steady state by having a gap of at least 10 minutes between
different measurements. To further minimize the error, it is necessary to repeat each experiment by 
interchanging the thermocouples. This cancels out all sources of linear error. Each such
experiment set is repeated 10 times. We report the mean values. Because of 
mechanical constraints, we were not able to attach more than two thermocouples at the
same time (see Figure~\ref{fig:position}).

We also report the CPU power as measured by the Windows CPUID utility. This is
the temperature at the center of the die~\cite{intel-therm}.
As a benchmark, we use a simple script that repeatedly performs calculations using
the standard Windows calculator application.

\begin{table*}[!htb]
\begin{center}
\begin{tabular} {||l||c|c|c||c|c|c||c|c|c||c|c|c||}
\hline
\hline
   & \multicolumn{3}{c||}{Left-Bottom (TCB and TCL)} 
   & \multicolumn{3}{|c||}{Right-Top(TCR and TCT)}  
   & \multicolumn{3}{|c||}{Top-Bottom(TCT and TCB)}  
   & \multicolumn{3}{c||}{Left-Right(TCL and TCR)}  \\
\hline
Operation & CPU  &  TCB & (TCB 
		  & CPU  &  TCT & (TCT  
          & CPU  &  TCT & (TCT
          & CPU  &  TCR & (TCR \\
          & ($^\circ$C) & ($^\circ$C) & -TCL)($^\circ$C)   
          & ($^\circ$C) & ($^\circ$C)  &  -TCR)($^\circ$C)    
          & ($^\circ$C) & ($^\circ$C)  &  -TCB)($^\circ$C)    
          & ($^\circ$C) & ($^\circ$C)  &  -TCL)($^\circ$C)    \\
\hline
Power Off  			&   &  23.90 &  0.15    &      & 20.90  &  0.50 &       & 24.75 & 0.25   &      & 25.80  &  0.40 \\
\hline
10 mins later  		& 57.5 &  39.85 &  4.05 & 52.0 & 35.95  &  3.10 & 52.5  & 39.65  & 7     & 43.5 & 33.50  &  1.35 \\
\hline
10 mins after  		& 89.0 &  58.50 & 6.80  & 90.0 & 55.80  &  6.35 & 88.0  & 62.55 & 14.35  & 74.5 & 56.05  & -1.15   \\
calculator on  &  &   &   &  &   &   &  &   &  &  &   &  \\
\hline
10 mins after 		& 58.5 &  41.25 & 4.20  & 51.5 & 37.85  &  4.45 & 53.5  & 41.40  & 9.85  & 43.5 & 33.45  &  2.00 \\
calculator off  &  &   &   &  &   &   &  &   &  &  &   &  \\
\hline
\hline
\end{tabular}
\end{center}
\caption{Temperatures of Points on the Surface of the Heat Spreader \label{tab:spreader}}
\end{table*}

Table~\ref{tab:spreader} shows the collected data at four time instants -- power off, 10 minutes later, 10 minutes after
starting the benchmark, and 10 minutes after shutting it down. We report four sets of readings. 

The die temperature varies from 51\celsius to 89\celsius. The spreader temperature at the hottest
point (TCT) varies from 21\celsius to 62.55\celsius. As a sanity check we correlate the temperature values with the
layout of the Pentium 4 processor~\cite{diephoto}. TCT is close to the scheduler and trace-cache.
In comparison TCB is the coolest because it abuts the L2 cache. TCL and
TCR are closest to the fetch/decode logic, and floating point units respectively. They show a moderate amount
of activity for our benchmark. 

The main take-away point in this experiment is that a large temperature variation exists across the surface of the spreader.
For example, the difference between TCT and TCB reaches 14.35\celsius.  The values at TCT and TCL differ by about 7\celsius,
and both TCR and TCL are warmer than TCB by about 6\celsius. This experiment gives us an indication of the degree
of the temperature gradients on the surface of the spreader. However, to get a more exact picture, we need to do a more
intrusive experiment. 

\subsection{Experiment II - IR Camera based Measurement}
To get accurate and extensive temperature profiles, we decided to use an IR camera that can produce a detailed temperature
profile of the surface of the spreader. A similar approach has been used by Martinez et. al.~\cite{martinez} to capture
the temperature profile of a die. The authors in this paper collect their data by removing the spreader and heat sink. They
use an IR transparent oil based heat sink instead. Note that it is necessary to use some heat removal mechanism. Otherwise, the temperature
of the die will increase to unacceptable levels, and the processor will shut itself down. We are planning to create
such kind of a setup in the future. However, we observe that in such a setup we will not get an accurate picture of
the temperature dissipation of a die and the thermal profile of the spreader because the nature of heat transfer
is different. The latest version of the popular thermal modeling tool HotSpot 5 takes this into
cognizance. Additionally, secondary heat
transfer paths, especially through the ball grid arrays, become important in this case.

\begin{figure}[!htb]
\includegraphics[width=0.80\columnwidth]{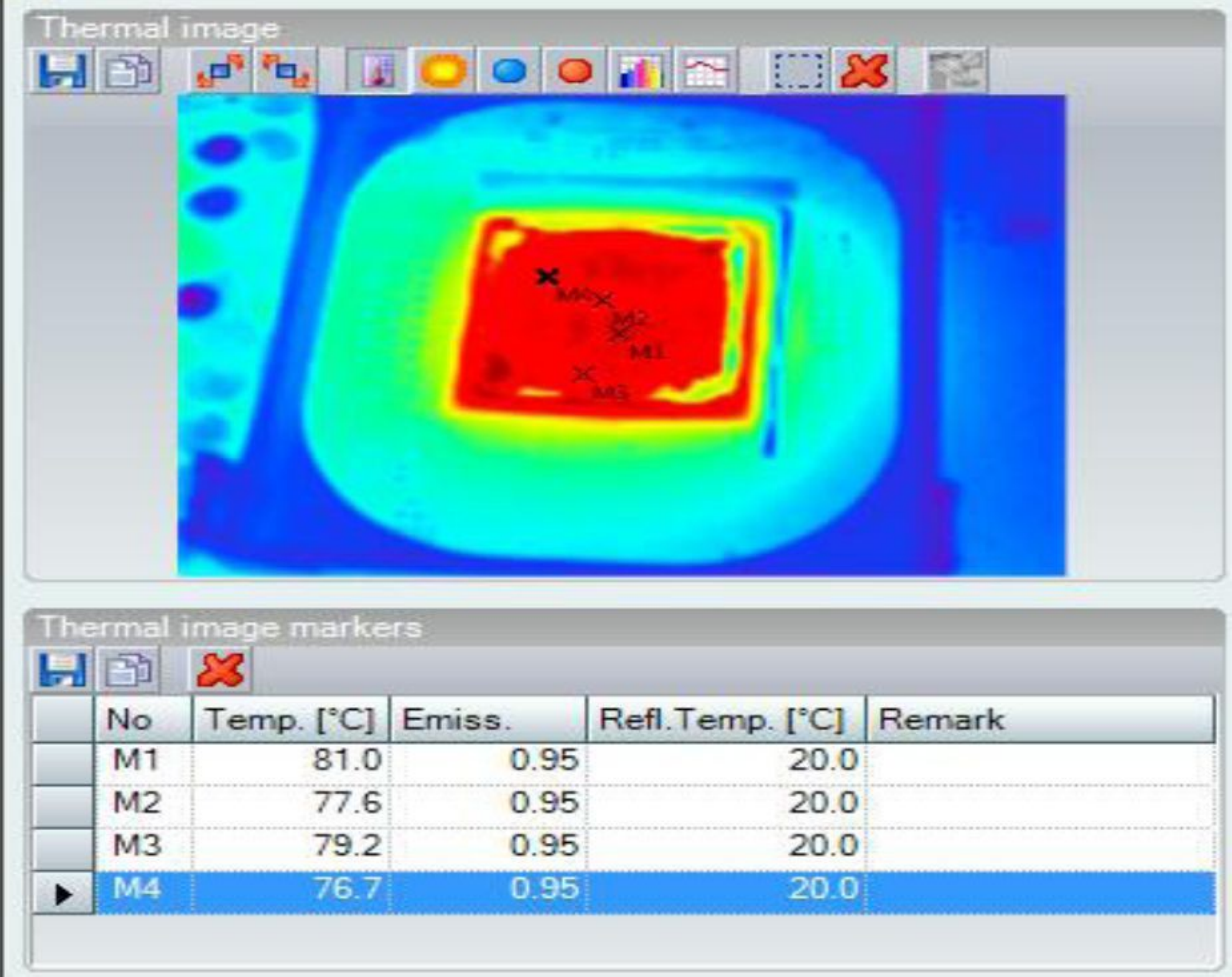}
\caption{IR Photograph \label{fig:ir}}
\end{figure}

We unsuccessfully try another approach. Our intuition was to remove the heat sink during regular operation and quickly
take an IR photograph of the die. There will be an intermittent delay of less than a few seconds. However, we hoped
to possibly compensate for the error by trying to back calculate the original temperature profile using standard 
results for radiative and convective heat transfer. We use a Testo 875, 9 Hz IR camera for this purpose. We initially
underestimate the temperature of the spreader greatly. This is possibly because of the low emissivity of the heat
spreader. Consequently, to increase the emissivity of the spreader, we coat it with Halnziye HY 610 thermal paste 
 that has an emissivity
of 0.95. The temperature values increase by about 15\celsius. The average temperature difference is about 10\celsius
across the die. However, our original aim of getting a detailed temperature profile was still not served because 
the thermal image was heavily dependent on the uniformity of the thermal paste. As shown in Figure~\ref{fig:ir} some
of the hottest(darkest) regions are towards TCB (L2 cache). This is not expected to be the case.

\begin{figure*}[!htb]
\begin{center}
\begin{tabular}{cc}

\begin{minipage}{0.5\textwidth}
\includegraphics[width=0.99\columnwidth]{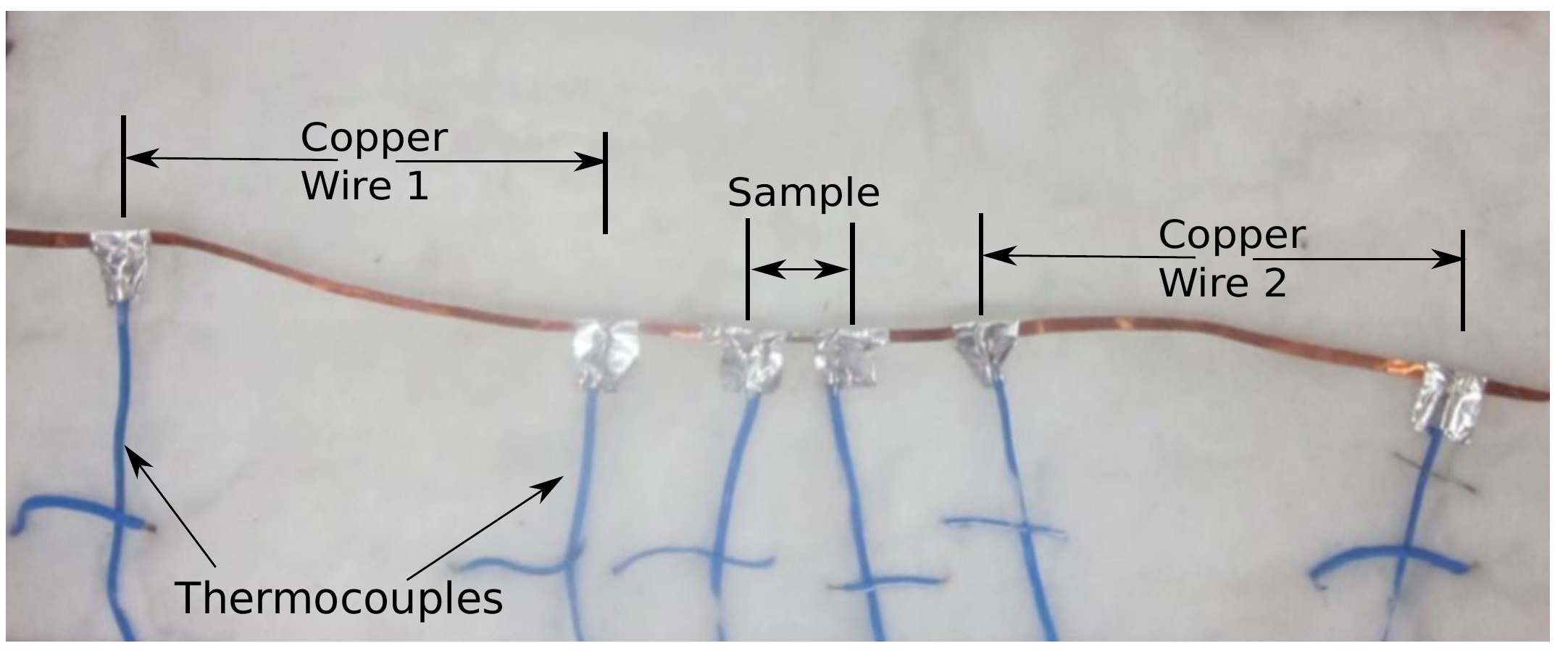}
\caption{Measurement of thermal conductance \label{fig:setupcomp}}
\end{minipage}

&

\begin{minipage}{0.5\textwidth}
\includegraphics[width=0.99\columnwidth]{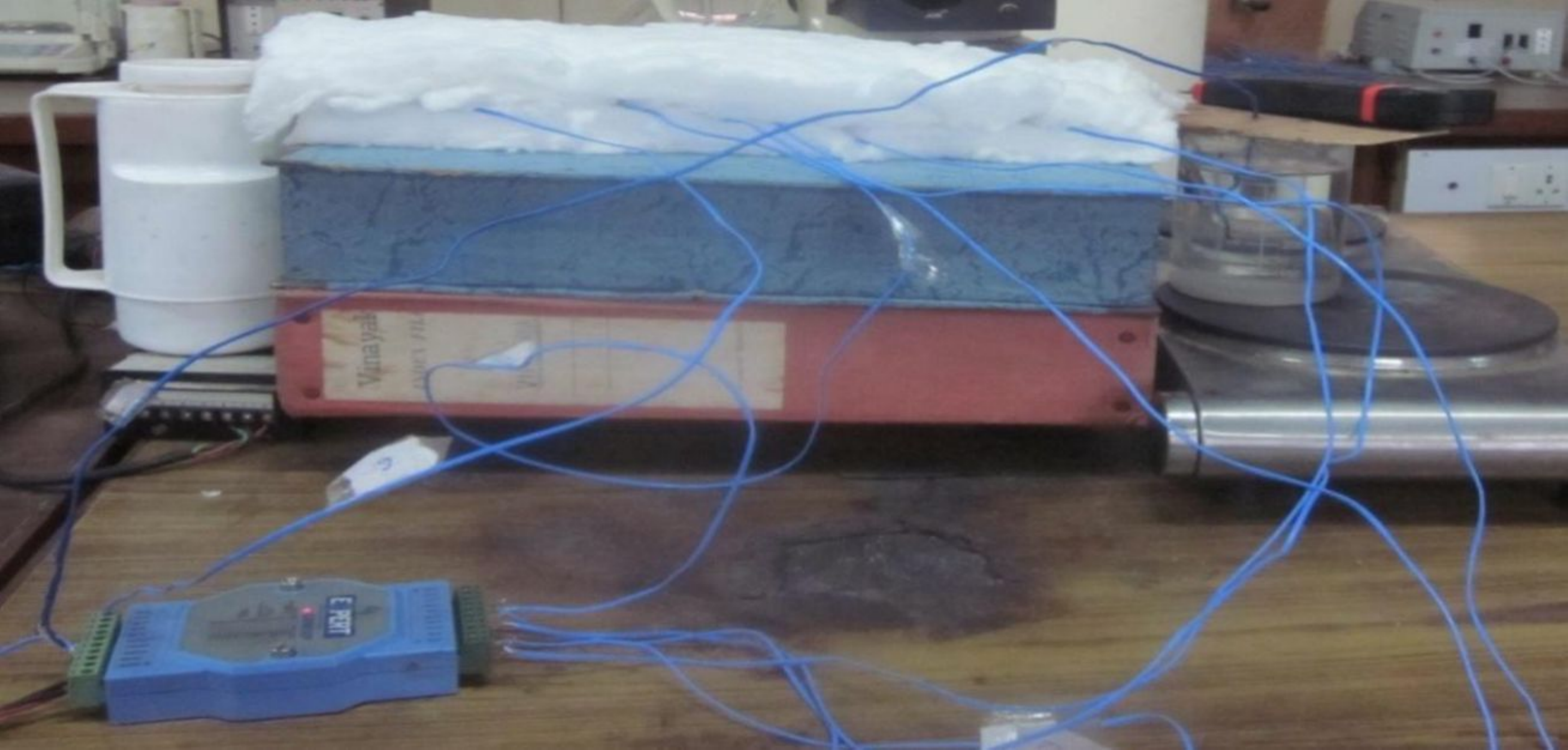}
\caption{The full measurement setup \label{fig:fullsetup}}
\end{minipage}
\\

\end{tabular}
\end{center}
\end{figure*}

\subsection{Experiment III - Measuring Thermal Conductivity}
We observe that Experiment II was inconclusive. It only reaffirmed the fact that a temperature differential exists
across the spreader. To get a better picture, we are in the process of setting up a detailed FEM simulation framework
that will be seeded by parameters obtained from our experiments. We describe a method to calculate the thermal
conductivity of the spreader material (nickel coated copper plate). The {\em thermal conductivity} of a material is defined as the
power that flows across a temperature gradient of 1\celsius in an object that has unit length  and unit cross-sectional
area. Conceptually, it is similar to electrical conductivity, and can be used to find the steady state distribution of
temperature. 

We use the comparative method. This method proposes to place the unknown sample (sliver of the spreader material) between
two samples (copper wires) with known thermal conductivity. Both the ends of this ensemble are set to constant temperatures
by dipping them in ice and boiling water respectively. The thermal conductivities are related by the following equation. 
\begin{equation}
\frac {\kappa_{c} \Delta T_{w1} A_{w1}}{L_{w1}} = \frac{\kappa_{sp} \Delta T_{sp} A_{sp}}{ L_{sp}} = \frac {\kappa_{c} \Delta T_{w2} A_{w2}}{L_{w2}} 
\label{eqn:comp}
\end{equation}

Here, $\kappa_{c}$ is the thermal conductivity of copper (400 W/m-K), and $\kappa_{sp}$ is the unknown thermal conductivity of the
spreader. $w_1$ refers to copper wire 1, $w_2$ refers to copper wire 2, $sp$ refers to the spreader sample, $A$ represents the cross-sectional
area, and $L$ represents the length of the wire. The temperature gradient, $\Delta T$, is measured using thermocouples. 
The intuition behind this equation is that there is a constant amount of heat flow in the assembly of wires. 
Note that Equation~\ref{eqn:comp} is over-constrained. We create two sets of equations -- (1) between the spreader and wire 1,
and (2) between the spreader and wire 2. We solve them separately and report the mean value of thermal conductivity. 

Figure~\ref{fig:setupcomp} shows the measurement setup with the two copper wires, spreader sample, and six thermocouples. 
Each adjacent pair of thermocouples measures the temperature difference across a homogeneous section of material. 
Note that Equation~\ref{eqn:comp} can be used only when there is exclusively conductive heat transfer through the
ensemble. We need to reduce convective and radiative heat transfer to the maximum extent possible. Consequently, we covered
the setup with thermally insulating glass wool. Lastly, we set the reference temperature at both ends using boiling
water and ice respectively. The entire setup is shown in Figure~\ref{fig:fullsetup}. We allowed upto 4-5 hours
for the readings to stabilize. The experimental procedure (10 repetitions, thermocouple interchange) is the same as that
mentioned in Section~\ref{sec:exp1}.

We obtain a thermal conductivity value of 369 W/m-K ($\pm 0.5$\celsius). For reference, the thermal conductivity of
copper is 400 W/m-K, and the thermal conductivity of nickel is 90.9 W/m-K. 

\section{Conclusion and Future Work}
In the course of this work, we get an in-vivo estimate of the nature of temperature gradients on the surface
of the heat spreader. We would like to extend our work to make more detailed and elaborate measurements. In
specific, we would like to drill very fine holes (diameter less than 100 nm) on the heat sink and measure
the temperatures at the center of the spreader also. Lastly, using empirical data collected from our studies, 
we wish to correlate our measurements with FEM based simulations. The final goal is four fold -- (1) Create
a corpus of empirically measured temperature data, (2) Propose accurate temperature simulation methodologies
for semi-conductor packages, and (3) Design new packaging technologies that are more thermally efficient, 
and lastly (4) Come up with new architectures that can leverage these advances in novel packaging technologies.


\end{document}